\documentclass[a4paper, twocolumn, 11pt, unpublished]{quantumarticle}
\pdfoutput=1
\usepackage[utf8]{inputenc}
\usepackage[english]{babel}
\usepackage[T1]{fontenc}
\usepackage{amsmath, mathtools, amsfonts, amsthm, amssymb}
\usepackage{hyperref}
\usepackage{nicefrac}
\usepackage{braket}
\usepackage{balance}
\DeclareMathOperator{\diag}{diag}
\usepackage[scr=boondox]{mathalpha}

\usepackage[linesnumbered,ruled,lined]{algorithm2e}

\makeatletter
\def\thmheadbrackets#1#2#3{%
  \thmname{#1}\thmnumber{\@ifnotempty{#1}{ }\@upn{#2}}%
  \thmnote{ {\the\thm@notefont(#3)}}}
\makeatother

\newtheoremstyle{brakets}
  {}
  {}
  {\itshape}
  {}
  {\bfseries}
  {.}
  { }
  {\thmheadbrackets{#1}{#2}{#3}}

\theoremstyle{brakets}
\newtheorem{theorem}{Theorem}
\newtheorem{lemma}{Lemma}

\newtheorem{assumption}{Assumption}

\newtheorem{remark}{Remark}
\newtheorem{corollary}{Corollary}
\newtheorem{problem}{Problem}
\newtheorem{conjecture}{Conjecture}

\title{$i$Trust: Trust-Region Optimisation with Ising Machines}

\author{Sayantan Pramanik}
\affiliation{Corporate Incubation, TATA Consultancy Services, India}
\affiliation{Robert Bosch Centre for Cyber-Physical Systems, Indian Institute of Science}
\email{sayantan.pramanik@tcs.com}
\author{Kaumudibikash Goswami}
\affiliation{QICI Quantum Information and Computation Initiative, The University of Hong Kong}
\author{Sourav Chatterjee}
\affiliation{Corporate Incubation, TATA Consultancy Services, India}
\author{M Girish Chandra}
\affiliation{TCS Research, TATA Consultancy Services, India}

\begin{document}

\newcommand{\J}{\boldsymbol{J}}
\newcommand{\h}{\boldsymbol{h}}
\newcommand{\s}{\boldsymbol{s}}
\newcommand{\sk}{\boldsymbol{s}^{(k)}}
\newcommand{\skp}{\boldsymbol{s}^{(k+1)}}
\newcommand{\sts}{\boldsymbol{s}_{(t)}^*}
\newcommand{\es}{E(\boldsymbol{s})}
\newcommand{\esk}{E(\boldsymbol{s}^{(k)})}
\newcommand{\eskp}{E(\boldsymbol{s}^{(k+1)})}
\newcommand{\ges}{\nabla E(\boldsymbol{s})}
\newcommand{\gesk}{\nabla E(\boldsymbol{s}^{(k)})}
\newcommand{\nk}{\boldsymbol{\zeta}^{(k)}}
\newcommand{\C}{\mathscr{C}}
\newcommand{\g}{\mathscr{g}}
\newcommand{\Ct}{\mathscr{C}_t}
\newcommand{\B}{\mathscr{B}}
\newcommand{\Bt}{\mathscr{B}_t}
\newcommand{\x}{\boldsymbol{\theta}}
\newcommand{\xt}{\boldsymbol{\theta}^{(t)}}
\newcommand{\xtp}{\boldsymbol{\theta}^{(t+1)}}
\newcommand{\p}{\boldsymbol{p}}
\newcommand{\pt}{\boldsymbol{p}^{(t)}}
\newcommand{\pts}{\boldsymbol{p}_{(t)}^*}
\newcommand{\gx}{\nabla f(\x)}
\newcommand{\gxt}{\nabla f(\xt)}
\newcommand{\Hess}{\boldsymbol{H}}
\newcommand{\gmk}{\boldsymbol{g}_{\beta_k}^{(k)}}


\twocolumn[
\begin{@twocolumnfalse}
\maketitle
\begin{abstract}
\textbf{In this work, we present a heretofore unseen application of Ising machines to perform trust region-based optimisation with box constraints. This is done by considering a specific form of opto-electronic oscillator-based coherent Ising machines with clipped transfer functions, and proposing appropriate modifications to facilitate trust-region optimisation. The enhancements include the inclusion of non-symmetric coupling and linear terms, modulation of noise, and compatibility with convex-projections to improve its convergence. The convergence of the modified Ising machine has been shown under the reasonable assumptions of convexity or invexity. 
The mathematical structures of the modified Ising machine and trust-region methods have been exploited to design a new trust-region method to effectively solve unconstrained optimisation problems in many scenarios, such as machine learning and optimisation of parameters in variational quantum algorithms. Hence, the proposition is useful for both classical and quantum-classical hybrid scenarios. 
Finally, the convergence of the Ising machine-based trust-region method, has also been proven analytically, establishing the feasibility of the technique.} 
\end{abstract}
\end{@twocolumnfalse}]


\section{Introduction}
Increasing demand for solving large-scale optimisation problems has recently motivated the development of non-von-Neumann-computer-based methods~\cite{Basu2019,Basu2020}, including Ising models~\cite{Barahona1982}, which rely on the physical dynamics of a system. Ising models have traditionally been used to solve $\textsc{NP}$-hard 
 combinatorial optimisation problems~\cite{kalinin2023analog, McMahon2016} by exploiting the adiabatic evolution of a physical system. Specifically, such problems are solved by mapping them to the ground-state search problem of the Ising model~\cite{Lucas2014}, where the ground state encodes its optimal solution. Among various methods of realizing an Ising model of coupled artificial spins\,\cite{Kim2010,Wang2013,Barends2016}, an important approach is to utilise opto-electronic-oscillators (OEOs) for building a coherent Ising machine (CIM)~\cite{pmim,Prabhakar2022}. A CIM implements a network of artificial spins with bistable coherent optical states for mapping the optimisation problems to the ground state of the Ising model~\cite{Lucas2014, pmim}. The OEO-based CIM approach particularly stands out for its cost-effectiveness, ambient operation, and scope for miniaturization~\cite{Prabhakar2022}. Being inherently gain dissipative, it naturally approaches the optimal solution~\cite{pmim,Leleu2017}.

 In this work, we present a new application of OEO CIMs to unconstrained optimisation. This is the first time we have analytically proven the viability of Ising machines to perform trust-region-based optimization \cite{Nocedal, Sorensen, MnS} and refer to the technique as $i$Trust. The main advantage of this method stems from the avoidance of matrix-inversion, along with the other aforementioned benefits of OEO-CIMs.
 This opens up a new avenue of applications where the Ising machines may be used to optimise the parameters in arbitrary objective functions, with an important example being the objective (loss/reward/penalty) functions of machine learning (ML) \cite{sml, usl, rl}, quantum ML (QML) \cite{Benedetti2019, qml1, qml2}, and quantum-inspired ML (QiML) \cite{huynh2023quantuminspired} models and variational quantum algorithms (VQAs) \cite{Cerezo2021}.
 Hence, $i$Trust finds applicability in both classical and quantum-classical hybrid computing.  More generally, the optimisation of any parametrised, unconstrained objective function $f: \mathbb{R}^n \rightarrow \mathbb{R}$ is within the purview of $i$Trust. We denote the parameters of the objective function $f(\cdot)$ with the vector $\x \in \mathbb{R}^n$. For completeness, the overarching problem that we attempt to solve using $i$Trust is:
\begin{problem}\label{prob:f}
    \begin{equation}
        \min_{\x \in \mathbb{R}^n} f(\x),
    \end{equation}
\end{problem}
\!\!\!\!\!\!with the aim of finding a point $\x^*$ which satisfies second-order optimality conditions \cite{Nocedal}, under the following generic assumption \cite{Nocedal}: 
\begin{assumption}\label{as:bounded}
    If $\x^{(0)}$ is the starting point of an iterative algorithm, then the function $f(\cdot)$ is bounded below on the level set $\mathscr{S} = \{\x \, \vert \, f(\x) \leq f(\x^{(0)}) \}$ by some value $f^*$, such that $f^* \leq f(\x) \; \forall \; \x \in \mathscr{S}$. Further, $f$ is twice continuously differentiable on $\mathscr{S}$.
\end{assumption}

\section{Contributions}
Before proceeding with the rest of the paper, we summarise our major results and contributions made, along with the specific sections, theorems and corollaries. 

An overview of trust-region methods has been presented in Section \ref{sec:trm}, followed by a detailed description of $i$Trust in Section \ref{sec:itrust}. Algorithm \ref{alg:itrust} provides a comprehensive view of $i$Trust, while its convergence to second-order optimal points has been presented and proven in Theorem \ref{th:itrust}. The use of Ising machines for trust-region optimisation has been enabled by the proposal of appropriate modifications in Section \ref{sec:eecim}, and comparisons of this modified architecture with noisy projected gradient-descent has also been made. Finally, the convergence of the modified Ising machine has been shown for convex functions (Assumption \ref{as:conv}) with bounded gradients (Assumption \ref{as:bound_grad}), and smooth (Assumption \ref{as:lip}), locally invex (Assumption \ref{as:pl}) objective functions. For the convex case, the Ising machine converges to an $\varepsilon$-suboptimal solution with fixed step-sizes in $\mathcal{O}(\nicefrac{1}{\varepsilon^2})$ iterations (Theorem \ref{th:conv} and Corollary \ref{cor:conv}), and improved results have been presented in Theorem \ref{th:dim_conv} with decreasing steps. For invex objective functions, convergence to $\varepsilon$-suboptimal points has been shown in $\mathcal{O}(\ln(\nicefrac{1}{\varepsilon}))$ iterations with fixed step-sizes in Theorem \ref{th:pl_conv} and Corollary \ref{cor:pl_iter}.


\section{Enhancing the Ising Machine}\label{sec:ecim}
In this section, we focus on introducing the type of Ising machines under consideration, and the modifications made to enable trust-region optimisation. Specifically, we opt for the poor man's coherent Ising machine \cite{pmim} with clipped transfer functions, as suggested in \cite{clipped}. This form of the Ising machine, which we henceforth refer to as the Economical Coherent Ising Machine (ECIM), attempts to iteratively find the solution to the following constrained optimisation problem\footnote{The existence of two different objective functions in Problems \ref{prob:f} and \ref{prob:ising} may be confusing to the readers. We have tried our best to avoid this confusion by explicitly stating the specific function being referred to. In cases where the function has not been named, the choice should be clear from the context of the discussion.}:
\begin{problem}\label{prob:ising}
    \begin{equation}\label{eq:obj}
        \min_{\s \in [-\Delta, \Delta]^n}\left(\es \overset{\Delta}{=} \frac{1}{2}\braket{\s, \J \s} + \braket{\h,\s}\right)
    \end{equation}
\end{problem}
\!\!\!\!\!\!with $\h = \boldsymbol{0}$ and $\Delta = \nicefrac{1}{2}$, where the \textit{relaxed} decision variables are updated at each iteration $k$ as \cite{clipped}:
\begin{equation}\label{eq:pmim}
    \skp_i = \begin{cases}
                \alpha \sk_i \! - \beta \sum_j \J_{ij}\sk_j \! + \nk_i, \lvert \sk_i \rvert \leq 0.4\\
                0, \text{ otherwise}.
\end{cases}
\end{equation}
Here, $\nk \sim \mathcal{N}(\boldsymbol{0}, \sigma^2 \boldsymbol{I})$ is a random Gaussian noise; while $\alpha$, $\beta$, and the noise-variance $\sigma^2$, are hyperparameters. The noise helps one escape from unwanted local minima, maxima, and saddle points \cite{Jordan, gradient_noise, Ising}. The reason for the choice of this Ising machine, which will become clearer in Section \ref{sec:itrust}, is the ease with which it can be adapted to the requirements for trust-region optimisation. This stems from the use of a transfer function which is mostly linear in nature in its operating region. The non-linearity may be viewed as \textit{projection} back to the linear, feasible region. Further, the \textit{size} of the linear region can be easily manipulated by controlling the clipping threshold.


\subsection{Enhancements Made}\label{sec:eecim}

It was demonstrated in \cite{Ising} that a similar form of the ECIM, with trigonometric nonlinearity, is equivalent to noisy gradient descent by fixing $\alpha = 1$, and conflating $\beta$ with the step-size. The authors further enhanced it to include non-symmetric coupling matrices and external field terms without resorting to ancillary spins \cite{Ising, ancilla}. The convergence of the Ising machine was shown to improve with diminishing step-sizes $\beta_k$, and through modulation of the noise with the corresponding step-size at that iteration.

Inspired by \cite{Ising}, we replace the term $\J \sk$ in equation \eqref{eq:pmim} with $\gesk$, where $\ges = \nicefrac{(\J+\J^\top) \s}{2} + \h$ is the gradient of the unconstrained objective $\es$ with respect to $\s$. The presence of the linear term $\braket{\h,\s}$ is imperative for trust-region based optimisation, as discussed in detail in Section \ref{sec:itrust}. Next, we modulate the operating region of the ECIM by setting the clipping voltage to $\pm \Delta$, with $\Delta > 0$. This constrains the feasible region of the ECIM to a box in $\mathbb{R}^n$ that we denote by $\C$. Interestingly, the constraint $\s \in \C$ is equivalent to $\lvert\lvert \s \rvert\rvert_\infty \leq \Delta$. Finally, we modify the update equation of the ECIM such that it is consistent with the definition of projection onto the closed set $\mathcal{C} \subset \mathbb{R}^n$. The projection operator $\Pi_\mathcal{C}(\cdot)$ onto $\mathcal{C}$ is defined as \cite{PJ}:
\begin{equation}
    \Pi_\mathcal{C}(\boldsymbol{z}) \coloneqq {\underset {\boldsymbol{x\in \mathcal{C}}}{\arg\min}} \; \lvert \lvert \boldsymbol{x}-\boldsymbol{z} \rvert \rvert_2.
\end{equation}
As an aside, we reproduce the following important properties of projection onto convex sets from \cite{PJ}, which will play an important role in the upcoming analysis of the ECIM's convergence.
\begin{lemma}\label{lem:proj}
    If the set $\mathcal{C} \subset \mathbb{R}^n$ is convex, and $\boldsymbol{z} \in \mathbb{R}^n$, then $\forall \, \boldsymbol{x} \in \mathcal{C}$:
    \begin{equation}\label{eq:proj_ip}
        \braket{\boldsymbol{x}-\Pi_\mathcal{C}(\boldsymbol{z}), \boldsymbol{z}-\Pi_\mathcal{C}(\boldsymbol{z})} \leq 0,
    \end{equation}
    and,
    \begin{equation}\label{eq:proj_contra}
        \lvert\lvert \Pi_\mathcal{C}(\boldsymbol{z}) - \boldsymbol{x} \rvert\rvert_2 \leq \lvert\lvert \boldsymbol{z} - \boldsymbol{x} \rvert\rvert_2.
    \end{equation}
\end{lemma}
Equation \eqref{eq:proj_contra} in the above lemma is referred to as the \textit{contraction} property of convex projections, and may be intuitively interpreted as the action of the projection operator diminishing or contracting the distance between the points $\boldsymbol{x} \in \mathcal{C}$ and $\boldsymbol{z} \in \mathbb{R}^n$. Since $\boldsymbol{x}$ is already in $\mathcal{C}$, the lemma may be written as $\lvert\lvert \Pi_\mathcal{C}(\boldsymbol{z}) - \Pi_\mathcal{C}(\boldsymbol{x})\rvert\rvert_2 \leq \lvert\lvert \boldsymbol{z} - \boldsymbol{x} \rvert\rvert_2$.

For the convex box $\C$, the projection operator $\Pi_{\C}(\cdot)$ is given by:
\begin{equation}
    \Pi_{\C}(\boldsymbol{z_i}) = \begin{cases}
                                    \boldsymbol{z}_i, \text{\;\;if $\lvert\boldsymbol{z}_i\rvert \leq \Delta$}\\
                                    \Delta, \text{\;\;if $\lvert\boldsymbol{z}_i\rvert > \Delta$.}
    \end{cases}
\end{equation}
Consequently, the update equation for the modified ECIM can be comprehensively stated as:
\begin{equation}\label{eq:ecim}
    \skp = \Pi_{\C}\left(\sk - \beta_k\left(\gesk - \nk\right) \right).
\end{equation}

As an extension to \cite{Ising}, similarities may be drawn between equation \eqref{eq:ecim} and noisy projected gradient descent  \cite{PJ}.


\subsection{Analysis of Convergence}
Having discussed the requisite changes to the ECIM, we now focus our attention towards analytically examining its performance on two classes of functions: one where $E(\cdot)$ is \textit{convex}, and the other where $E(\cdot)$ is \textit{smooth} and \textit{locally invex} on some domain. This domain is the constraint set $\C$ of Problem \ref{prob:ising}. However, functions of these categories are known to lack saddle points, and in the constrained case, the maxima mostly lie on the boundary of the feasible region. As a result, any stationary point is a global minimum on $\C$. As the primary utility of the injected-noise was to avoid saddle points and (local) maxima, its necessity is eliminated for the functions under consideration. Hence, the white noise is removed from the ECIM by setting $\sigma^2=0$. This also removes any stochasticity from the performance of the Ising machine, unlike the work presented in \cite{Ising}. 

As a quantifiable measure of convergence, we track the \textit{suboptimality gap} over the iterations of the ECIM, and report explicit rates of convergence with both constant and diminishing step-sizes. When the function is convex, we define the suboptimality gap $\g$ to be a measure of the difference between the best observed function value over all the iterations and the optimal value $E^*$ in the feasible region. While in the case of smooth, invex functions, it is defined slightly differently as the difference between the objective value at the $k^{\text{th}}$ iteration and $E^*$. We show that in both cases, the ECIM is able to get arbitrarily close to the optimal objective value. The analyses for locally convex and invex functions may be found in the upcoming subsections \ref{sec:conv} and \ref{sec:pl}, respectively. The techniques used to prove the ensuing theorems and corollaries is similar to those in \cite{GD, PJ}, and by extension, to the references therein.

\subsubsection{Convex Functions}\label{sec:conv}
In this subsection, we impose convexity on the objective function $E$, as formalised below:
\begin{assumption}[Convexity of $E$]\label{as:conv}
    The function $E(\cdot)$ is assumed to be convex, i.e.,
    \begin{enumerate}
        \item for every $\s$, $\mathscr{\boldsymbol{z}}$ $\in \mathbb{R}^n$, $E(\mathscr{\boldsymbol{z}})-\es \geq \braket{\ges, \mathscr{\boldsymbol{z}}-\s}$,
        \item $\nicefrac{(\J+\J^\top)}{2}$ is positive semidefinite;
    \end{enumerate}
    with the converse is also being true.
\end{assumption}

\!\!\!\!\!\!Additionally, we require the gradients of the objective function to be bounded:
\begin{assumption}[Bounded Gradients]\label{as:bound_grad}
    We also assume that the gradients of $\es$ with respect to $\s$ are bounded, i.e., there exists a non-negative $G < \infty$ such that $\forall \, \s \in \C$:
    \begin{equation}
        \lvert\lvert \ges \rvert\rvert_2 \leq G.
    \end{equation}
\end{assumption}

Equipped with Assumptions \ref{as:conv} and \ref{as:bound_grad}, the convergence of the ECIM for convex objective functions may be derived and stated as below:
\begin{theorem}[Convergence with Fixed Step-Sizes]\label{th:conv}
    If Assumptions \ref{as:conv} and \ref{as:bound_grad} hold true, and the update in \eqref{eq:ecim} is run for $K \in \mathbb{N}$ iterations with a fixed step-size $\beta$ to generate a sequence $(\sk)_{k \leq K}$, then for every $K \geq 1$, we have:
    \begin{equation}
        \g \leq \frac{1}{2}\left( \frac{\lvert\lvert \s^{(0)}-\s^* \rvert\rvert^2_2}{\beta K} + \beta G^2 \right),
    \end{equation}
    where $\g = \min(\esk) - E^*$, $E^* = \min_{\s \in [-\Delta, \Delta]^n} \es$, and $\s^*$ is a minimiser of $\es$ in $\C$.
\end{theorem}

\begin{corollary}[Iteration Complexity with Fixed Step-Sizes]\label{cor:conv}
    In the setting of Theorem \ref{th:conv}, if the ECIM is run for $K \in \mathbb{N}$ iterations with $\beta = \nicefrac{\beta_0}{\sqrt{K}}$, where $\beta_0 = \nicefrac{\lvert\lvert \s^{(0)}-\s^* \rvert\rvert_2}{G}$, then for every $\varepsilon > 0$ it can be guaranteed that $\left(\min(\esk) - E^*\right) \leq \varepsilon$. Moreover,
    \begin{equation}
        K \geq \left(4\lvert\lvert \s^{(0)}-\s^* \rvert\rvert_2^2 G^2\right)\frac{1}{\varepsilon^2}.
    \end{equation}
\end{corollary}
It may be inferred from Theorem \ref{th:conv} and Corollary \ref{cor:conv} that the suboptimality gap obtained by the ECIM may be bounded above by an arbitrarily small, positive quantity $\varepsilon$, i.e., it is possible to get arbitrarily close to $E^*$ with the right choice of step-size $\beta$. Such a point is referred to as an $\varepsilon$-suboptimal point. Further, the ECIM is guaranteed to find such a point in $\mathcal{O}\left(\nicefrac{1}{\varepsilon^2}\right)$ iterations \cite{GD}.

A major drawback of the above result is that the step-size employed is a \textit{fixed-horizon} one \cite{PJ}, i.e., it is dependent on the total number of iterations $K$, which needs to be predetermined. Also, the value of $\beta$ is dependent on $\lvert\lvert \s^{(0)}-\s^* \rvert\rvert_2$ and $G$, the values of which are, in general, unknown. This leads to difficulties in implementation. This issue may be easily bypassed through the use of decreasing step-sizes. We resort to a sequence of step-sizes that conform to the conditions in equation \eqref{eq:beta}, and show in Theorem \ref{th:dim_conv} that convergence may still be attained. The readers may also note that these conditions on the step-sizes are prevalent in machine learning literature and practice.

\begin{theorem}[Convergence with decreasing step-sizes]\label{th:dim_conv}
    In the setting of Theorem \ref{th:conv}, if the ECIM is instead run with decreasing sequence of step-sizes $(\beta_k)$ such that:
    \begin{equation}\label{eq:beta}
       \sum_{k=0}^{\infty}\beta_k = \infty \text{  and  } \sum_{k=0}^{\infty}\beta_k^2 < \infty, 
    \end{equation} then 
    \begin{equation}
        \lim_{K \rightarrow \infty} \left( E(\Bar{\s}^{(K)}) - E^* \right) = 0,
    \end{equation}
    where $\Bar{\s}^{(K)} = \frac{1}{\sum_{k=0}^{K-1}\beta_k}\sum_{k=0}^{K-1}\beta_k\sk$.
\end{theorem}

\subsubsection{Invex Functions}\label{sec:pl}
In this subsection, we assume the objective function to be smooth and locally invex. To start with, we define smoothness in the assumption below, followed by a discussion on invexity:
\begin{assumption}[Lipschitz Smoothness]\label{as:lip}
    The function $E(\cdot)$ is assumed to be L-Lipschitz smooth, i.e., there exists an $L > 0$ such that $\forall \; \s, \boldsymbol{\mathscr{s}} \in \mathbb{R}^n$:
    \begin{equation}
        \lvert\lvert \nabla E(\s) - \nabla E(\boldsymbol{\mathscr{s}}) \rvert\rvert_2 \leq L \lvert\lvert \s - \boldsymbol{\mathscr{s}} \rvert\rvert_2.
    \end{equation}
\end{assumption}

While invexity may not be as popular a condition as convexity, it is definitely more general. For instance, invex functions have been shown to include neural networks with ReLU activations and quadratic losses where convexity cannot be assumed \cite{GD, 6inGD}. Further, \cite{Schmidt} argues and proves that among Lipschitz-smooth functions such as \textit{strongly convex}, \textit{essentially strongly convex}, \textit{weakly strongly convex}, and functions obeying the \textit{restricted secant inequality}, invex functions entail the weakest assumptions. A more detailed exposition on the relations and implications between function-classes may be found in Theorem 2 of \cite{Schmidt}. Further, it is known that invex functions obey the Polyak-$\L$ojasiewicz (P\L) inequality. We, however, require the objective function to be invex \textit{locally} on the constraint set $\C$. Consequently, the $E$ follows the local P\L \, inequality, as mentioned in the assumption below: 
\begin{assumption}[Local P\L \, inequality \cite{Schmidt}]\label{as:pl}
    The function $E(\cdot)$ obeys the P\L \, inequality on $\C$ such that for some $\mu > 0$ and for all $\s \in \C$,
    \begin{equation}\label{eq:pl}
        \lvert\lvert \ges \rvert\rvert_2^2 \geq 2\mu (\es - E^*).
    \end{equation}
\end{assumption}

Proceeding further, we recall the ECIM's update rule from equation \eqref{eq:ecim}, and define the following \textit{gradient mapping} vector $\gmk$ \cite{ubc} as:
\begin{equation}\label{eq:gmk}
    \gmk = \frac{1}{\beta_k}(\sk - \skp).
\end{equation}
It is interesting to note that for unconstrained problems, $\gmk$ reduces to the noisy gradient update $\gesk - \nk$. Thus, the gradient mapping vector may be viewed as a counterpart of the gradient vector for constrained optimisation problems. In the following lemma, we state two important relations between $\gmk$ and $\gesk$:
\begin{lemma}
    For the iterative update rule in \eqref{eq:ecim}, and the convex set $\C$, if the vector $\gmk$ is defined as in equation \eqref{eq:gmk}, then:
    \begin{equation}\label{eq:gmk1}
        \braket{\gesk, \skp - \sk} \leq \beta_k \lvert\lvert \gmk \rvert\rvert_2^2,
    \end{equation}
    and,
    \begin{equation}\label{eq:gmk2}
        \lvert\lvert \gmk \rvert\rvert_2^2 \leq \lvert\lvert \gesk \rvert\rvert_2^2.
    \end{equation}
\end{lemma}
Based on equations \eqref{eq:pl} and \eqref{eq:gmk2}, we conjecture the following P\L-like relation between $\gmk$ and the suboptimality gap at the $k^{\text{th}}$ iteration:
\begin{conjecture}\label{conj}
    If Assumption \ref{as:pl} is true, then there exists a $\mu_p$, $0 < \mu_p \leq \mu$, such that $\forall \, \s \in \C$ and $\forall \, \beta \in \mathbb{R}^+$:
    \begin{equation}
        \lvert\lvert \gmk \rvert\rvert_2^2 \geq 2\mu_p (\esk - E^*).
    \end{equation}
\end{conjecture}

\begin{theorem}[Convergence with Fixed Step-Sizes]\label{th:pl_conv}
    Considering Assumption \ref{as:lip} and Conjecture \ref{conj} to hold true, if $(\sk)_{k \leq K}$ is the sequence of iterates produced by equation \eqref{eq:ecim} over $K \in \mathbb{N}$ iterations with a constant step-size $0 < \beta \leq \nicefrac{1}{L}$, then:
    \begin{equation}
        E(\s^{(K)}) - E^* \leq (1-\beta \mu_p)^K(E(\s^{(0)})-E^*).
    \end{equation}
\end{theorem}

\begin{corollary}[Iteration Complexity with Constant Step-Sizes]\label{cor:pl_iter}
    In the setting of Theorem \ref{th:pl_conv}, for any $\varepsilon > 0$, it can be guaranteed that $\left(E(\s^{(K)}) - E^*\right) \leq \varepsilon$. Moreover,
    \begin{equation}
        K \geq \frac{L}{\mu_p}\ln\left( \frac{E(\s^{(0)})-E^*}{\varepsilon} \right).
    \end{equation}
\end{corollary}
\!\!\!\!\!\!It is clear from Theorem \ref{th:pl_conv} and Corollary \ref{cor:pl_iter} that when the objective function $E(\cdot)$ satisfies the P\L \, inequality, the ECIM is capable of converging \textit{linearly} to an $\varepsilon$-suboptimal solution in $\mathcal{O}\left(\ln\left(\nicefrac{1}{\varepsilon}\right)\right)$ iterations.

\subsection{Unification of Results}
If $\boldsymbol{\mathscr{s}}$ is the output of the ECIM, then we combine the results on the suboptimality gap obtained by the ECIM from Sections \ref{sec:conv} and \ref{sec:pl} in the following form, as suggested in \cite{MnS}:
\begin{equation}\label{eq:uni}
    -E(\boldsymbol{\mathscr{s}}) \geq c \lvert E(\s^*) \rvert,
\end{equation}
for some constant $c \in (0,1]$. This states that the final objective value reached by the ECIM is \textit{close} to the optimal value on $\C$, which is all that is required for trust-region methods to work, as detailed in the immediately-succeeding Section \ref{sec:trm}.

\section{Trust-Region Method}\label{sec:trm}
Before discussing how the ECIM may be used for trust-region based optimisation, we provide a brief overview of the latter in this section. Trust-region methods attempt to find an update to the current iterate $\xt$ by constructing a quadratic surrogate model that acts as a close approximation to the original objective function $f(\cdot)$ within a trust region. The trust region, often and conventionally, takes the shape of a ball\footnote{To avoid situations where the optimisation Problem \ref{prob:f} has a \textit{poor scaling} with respect to the decision variables $\x$, \textit{elliptical} trust regions may be employed by replacing the constraint of Problem \ref{prob:model} with:
\begin{equation}\label{eq:ellipse}
    \lvert\lvert\boldsymbol{D}\p\rvert\rvert_2\leq\delta,
\end{equation} 
where $\boldsymbol{D} = \diag(d_1, \dots, d_n)$ with $d_i \geq 0$. The elements $d_i$ are adjusted according to the \textit{sensitivity} of $f(\cdot)$ to $\x_i$: if $f(\cdot)$ varies highly with a small change in $\x_i$, then a large value of $d_i$ is used; and vice versa \cite{Nocedal}.} in $\mathbb{R}^n$ centered at $\xt$.
The surrogate model $m(\p)$ is inspired by the Taylor expansion of $f(\x+\p)$ around $\x$, which is given by:
\begin{equation}
    f(\x+\p) = f(\x) + \braket{\gx, \p} + \frac{1}{2}\braket{\p,\Hess(\boldsymbol{z})\p},
\end{equation}
where $\gx$ and $\Hess(\x)$ are the gradient and Hessian of $f$ at $\x$, respectively, and $\boldsymbol{z}$ is a convex combination of $\x$ and $\p$. Subsequently, the model is defined as:
\begin{equation}
    m_t(\p) = \braket{\gxt, \p} + \frac{1}{2}\braket{\p,\boldsymbol{B}^{(t)}\p}.
\end{equation}
Here, we explicitly set $\boldsymbol{B}^{(t)}$ to the (\textit{approximate}) Hessian $\Hess(\xt)$ to obtain a Newton-like method, but over a convex constraint-set \cite{Nocedal, Sorensen}. If the radius of the trust-region at iteration $t$ is $\delta_t$, then we represent the feasible set with $\Bt = \{\boldsymbol{z}\in\mathbb{R}^n \, \vert \, \lvert\lvert \boldsymbol{z}-\xt \rvert\rvert_2\leq\delta_t\}$. The update $\pts$ to $\xt$ is thus obtained by solving the following problem:
\begin{problem}\label{prob:model}
\begin{equation}\label{eq:tr_s_update}
    \underset{\p \in \Bt}{\min}\;m_t(\p).
\end{equation}
\end{problem}

\begin{remark}
    It is easy to note that since $f(\x+\p) \approx f(\x) + m_t(\p) + o(\lvert\lvert \p \rvert\rvert^2)$, $m_t(\pts)$ must be negative to obtain a decrease in value of the objective function $f$. 
\end{remark}
In practice, however, exact solutions to \eqref{eq:tr_s_update} are not necessary, and approximations to $\pts$ are used instead \cite{Nocedal}. The quality of approximation is quantified through the ratio $\rho_t$ of the actual and \textit{predicted} reductions in the function value \cite{Nocedal}:
\begin{equation}
    \rho_t = \frac{f(\xt+\pts)-f(\xt)}{m_t(\pts)}.
\end{equation}

The size of the trust-region is then varied based on the value of $\rho_t$ \cite{Nocedal, MnS, Sorensen} as:
\begin{enumerate}
    \item if $\rho_t$ is negative, the current iteration is rejected, and $\delta_t$ is reduced,
    \item if $\rho_t$ is positive, but significantly smaller than $1$, then the size of the trust-region is left unchanged,
    \item while if $\rho_t$ is close to $1$, then the trust-region is expanded.
\end{enumerate}
This process is repeated until desired convergence is achieved. We refer the reader to Algorithm 4.1 in \cite{Nocedal} for a comprehensive overview of trust-region based optimisation.

\section{$i$Trust}\label{sec:itrust}
A major disadvantage of using the method in Algorithm 3.14 of \cite{MnS} to find $\pts$ is the requirement for repeated Cholesky decomposition and inversion of the Hessian, both of which are in $\mathcal{O}(n^3)$. This becomes prohibitive for problems where $n$ is large, for instance machine learning models with millions of parameters. We aim to alleviate this problem by using the enhanced ECIM, described in Section \ref{sec:eecim} to find the update to $\xt$ at iteration $t$. We achieve this by exploiting the similarity in structure of Problems \ref{prob:ising} and \ref{prob:model} and using the modifies ECIM described in Section \ref{sec:ecim} to solve the trust-region subproblem. Specifically, at each iteration $t$, $\J$ is set to $\Hess(\xt)$, $\h$ to $\gxt$, and $\Delta$ to $\delta_t$. Here, the importance of the inclusion of linear terms in the Ising machine becomes clear, without which the gradient $\gesk$ could not have been provided to the ECIM without additional overheads in the form of ancillary spins \cite{Ising, ancilla}. We name this technique of using the ECIM for trust-region optimisation as $i$Trust. The workflow for $i$Trust has been portrayed in Algorithm \ref{alg:itrust}, which draws inspiration from, and is an amalgamation of, Algorithms 4.1 and 4.2 of \cite{Nocedal, MnS}, respectively. 
\begin{algorithm}
        \SetKwInOut{Input}{input}
        \Input{initial point $\x^{(0)} \in \mathbb{R}^n$; maximum trust-region radius $\delta_{\text{max}}>0$; initial radius $\delta_0 \in (0, \delta_{\text{max}}]$; thresholds on $\rho_t$: $0 < \mu < \eta <1$; radius-updation parameters $\gamma_1<1$ and $\gamma_2>1$; noise variance $\sigma^2$; sequence of step-sizes $(\beta_k)$; and number of iterations $T$ and $K$}
    \SetKwBlock{Beginn}{beginn}{ende}
    \Begin{
        \For{$t \in [T]$}{
                evaluate $\gxt$ and $\Hess(\xt)$\;
                $\J^{(t)} \leftarrow \Hess(\xt)$\;
                $\h^{(t)} \leftarrow \gxt$\;
                $\Delta_t \leftarrow \delta_t$\;
                initialise $\s^{(0)}$ randomly in $\C_t = [-\Delta_t, \Delta_t]^n$\;
                \For{$k \in [K]$}{
                sample $\nk \sim \mathcal{N}(\boldsymbol{0}, \sigma^2\boldsymbol{I})$\;
                $\skp = \Pi_{\C_t}\left(\sk - \beta_k\left(\nabla E_t(\sk) - \nk\right) \right)$\;
                }
                calculate $\rho_t = \frac{f(\xt + \s^{(K)})-f(\xt)}{E_t(\s^{(K)})}$\;
                \uIf{$\rho_t < \mu$} {
                    $\delta_{t+1} = \gamma_1 \delta_t$\;
                    \textbf{continue}\;
                }
                \Else {
                \uIf{$\rho_t > (1-\mu)$ and $\lvert\lvert \s^{(K)} \rvert\rvert_\infty=\delta_t$}{
                    $\delta_{t+1} = \min(\gamma_2\delta_t, \delta_{\text{max}})$\;
                    }
                \Else{$\delta_{t+1} = \delta_t$\;}
                }
                \uIf{$\rho_t > \eta$}{
                    $\xtp = \xt + \s^{(K)}$\;
                }
                \Else{$\xtp = \xt$\;}
        }
        \Return $\x^{(T)}$
    }
    \caption{$i\,$Trust}\label{alg:itrust}
\end{algorithm}

\begin{remark}
It is interesting to note that if the coupling matrix $\J^{(t)}$ is positive semidefinite at the iteration $t$, then as per the definition of convexity, the objective function of the trust-region subproblem is convex, satisfying Assumption \ref{as:conv}. Additionally, since the coupling matrix is equal to the Hessian $\Hess(\xt)$, this also implies that the objective function $f$ is convex in the region around $\xt$. Thus, in a convex region of the original problem, the analysis done in Section \ref{sec:conv} becomes applicable for the ECIM.
\end{remark}

Further, we distinguish between the minimisers of $E_t(\s)$ and $m_t(\p)$ on the sets $\C_t$ and $\B_t$ by denoting them with $\sts$ and $\pts$, respectively.

\begin{remark}
    We would like to emphasize that the box $\C_t$ and the ball $\mathscr{B}_t$ share a common centre $\xt$, and by design, the side-length of the box is set equal to the diameter of the ball at each iteration. Thus, the ball is contained completely within the box: $\mathscr{B}_t \subset \C_t$\footnote{Differential scaling with respect to different components of the decision variables may be handled by setting individual $\Delta_i$ for each coordinate $\x_i$ such that the elliptical trust-region from equation \eqref{eq:ellipse} lies within the box defined by the $\Delta_i$s.}. Now, since the objective function of the Problems \ref{prob:model} and \ref{prob:ising} are identical, and the constraint set of the former is contained in that of the latter, we have:
    \begin{equation}
        E_t(\sts) \leq m_t(\pts).
    \end{equation}
    This means that if the ECIM and the Algorithm 3.14 in \cite{MnS} can both reach near-optimal solutions of their respective optimisation problems, then the objective value obtained by the ECIM is guaranteed to be better. This results in a higher reduction in the value of $f(\x)$ at each iteration.
\end{remark}
\begin{theorem}[Convergence of $i$Trust]\label{th:itrust}
    Let Assumption \ref{as:bounded} be true, and let $(\xt)$ be the sequence of iterates generated by Algorithm \ref{alg:itrust} such that equation \eqref{eq:uni} is satisfied at each iteration. Then we have that:
    \begin{equation}
        \lim_{t \rightarrow \infty} \lvert\lvert \nabla f(\xt) \rvert\rvert_2 = 0.
    \end{equation}
    Moreover, if $\mathscr{S}$ is compact, the either Algorithm \ref{alg:itrust} terminates at a point $\x^{(T)} \in \mathscr{S}$ where $\nabla f(\x^{(T)}) = 0$ and $\Hess(\x^{(T)}) \succcurlyeq 0$; or $(\xt)$ has a limit point $\x^* \in \mathscr{S}$ such that $\nabla f(\x^*) = 0$ and $\Hess(\x^*) \succcurlyeq 0$.
\end{theorem}
The above theorem implies that $i$Trust is guaranteed to asymptotically converge to a stationary point of the function $f$. Furthermore, if the level set $\mathscr{S}$, defined in Assumption \ref{as:bounded} is closed and bounded, then the algorithm converges (or tends to converge) to a second-order optimal point in $\mathscr{S}$ \cite{Nocedal, MnS}.


\section{Conclusions and Outlook}
In this paper, we introduced $i$Trust, an algorithm that leverages Ising machines for trust-region based optimisation. In doing so, we proposed necessary modifications to the Ising machine, which we refer to as the Economical Coherent Ising Machine (ECIM). The feasibility and convergence of $i$Trust was also proven analytically. We look forward to the validation our theoretical results by experimenting extensively with the proposed algorithm. Possible future directions may include the investigation of the performance of the ECIM for other classes of objective functions besides convex and invex ones. Variants of $i$Trust can also be constructed that are compatible with natural gradient descent \cite{NGD, qng}, by replacing the Hessian with the Fisher Information Matrix. $i$Trust may be further augmented by zeroth order methods like SPSA \cite{spsa1} in scenarios where evaluation of the gradients, Hessian, and Fisher information matrix is computationally expensive \cite{QNSPSA}. Lastly, the advantages (or lack thereof) of the ECIM over projected gradient descent for the subproblem-minimisation can also be examined. We hope that this paper opens up new avenues of research in the analytical and empirical exploration of new applications of Ising machines.

\section*{Acknowledgement}
SP, SC, and MGC would like to express their gratitude towards Mr. Anil Sharma, Mr. Vidyut Navelkar (retd), Mr. C. V. Sridhar, Mr. Godfrey Mathais, and Dr. Anirban Mukherjee from the Corporate Incubation team at Tata Consultancy Services, and Mr. Manoj Nambiar from TCS Research for their unwavering support throughout the course of this work. SP would also like to thank Mr. Chaitanya Murti from the Indian Institute of Science for helpful discussions regarding Newton and Quasi-Newton methods.  KG is supported by the Hong
Kong Research Grant Council (RGC) through Grant No.
17307719 and 17307520.

\balance
\bibliography{refs}
\bibliographystyle{unsrt}

\end{document}